\documentclass[a4paper,10pt]{article}

\usepackage[utf8]{inputenc}
\usepackage[T1]{fontenc}

\usepackage{a4wide}

\usepackage{amsmath,amssymb}
\usepackage{amsthm}
\usepackage{enumerate} 

\usepackage{authblk}

\usepackage[dvipsnames,table,xcdraw,svgnames]{xcolor}
\definecolor{highlightNEW}{named}{black}
\definecolor{highlightCyan}{named}{Cyan}
\definecolor{highlightGreen}{named}{LightGreen}

\usepackage{hyperref}
\hypersetup{
    breaklinks=true,
    colorlinks=true,
    linkcolor=Navy,
    citecolor=Navy,
    urlcolor=Navy
}

\usepackage{graphicx}
\graphicspath{{./fig-paper/}{./fig/}}
\usepackage{epstopdf}
\usepackage[section]{placeins} % \FloatBarrier

\usepackage{booktabs}
\usepackage{multirow}

\usepackage[absolute,overlay,showboxes]{textpos}
\TPMargin{2mm}
\textblockrulecolour{Navy}

\usepackage[round]{natbib}
\bibpunct[, ]{(}{)}{;}{a}{}{,}

\usepackage{soul}

\soulregister\cite7%
\soulregister\citep7%
\soulregister\eqref7%
\soulregister\pageref7%
\soulregister\ref7%

\newcommand{\doi}[1]{DOI~\href{\detokenize{http://dx.doi.org/#1}}{\detokenize{#1}}}

\renewcommand{\d}{\,\mathrm{d}}

\newcommand{\E}{\mathbb{E}}
\newcommand{\F}{\mathcal{F}}

\newcommand{\Q}{Q}

\newcommand{\Var}{\operatorname{Var}}

\newcommand\abs[1]{\left|#1\right|} % absolute value

\newdimen\CdotAxis
\newcommand*{\CdotAux}[3]{%
  {%
    \settoheight\CdotAxis{$#2\vcenter{}$}%
    \sbox0{%
      \raisebox\CdotAxis{%
        \scalebox{#1}{%
          \raisebox{-\CdotAxis}{%
            $\mathsurround=0pt #2#3$%
          }%
        }%
      }%
    }%
    \dp0=0pt %
    \sbox2{$#2\bullet$}%
    \ifdim\ht2<\ht0 %
      \ht0=\ht2 %
    \fi
    \sbox2{$\mathsurround=0pt #2#3$}%
    \hbox to \wd2{\hss\usebox{0}\hss}%
  }%
}
\makeatletter
\def\mathcolor#1#{\@mathcolor{#1}}
\def\@mathcolor#1#2#3{%
  \protect\leavevmode
  \begingroup
    \color#1{#2}#3%
  \endgroup
}
\makeatother

\newcommand{\NEW}[1]{\mathcolor{highlightNEW}{#1}}
\let\oldalpha\alpha
\renewcommand{\alpha}{\mathcolor{highlightNEW}{\oldalpha}}
\newcommand{\ccode}[2]{\par
        \vspace*{3pt}
        {{\leftskip18pt\rightskip\leftskip
        \noindent{\it #1}\/: #2\par}}\par}
\newcommand{\keywords}[1]{\ccode{Keywords}{#1}}
\newcommand{\email}[1]{\href{mailto:#1}{#1}}

\def\received#1{Received~#1\par}
\def\revised#1{Revised~#1\par}

\usepackage[mathscr]{euscript}
\DeclareSymbolFont{rsfs}{U}{rsfs}{m}{n}
\DeclareSymbolFontAlphabet{\mathscrsfs}{rsfs}

\newcommand{\jpTitle}{Robustness and sensitivity analyses of rough Volterra stochastic volatility models}
\newcommand{\jpAuthors}{J. Matas and J. Posp\'{\i}\v{s}il}
\newcommand{\jpKeywords}{Volterra stochastic volatility; rough volatility; rough Bergomi model; robustness analysis; sensitivity analysis}
\newcommand{\jpMSC}{62F35; 62F40; 60G22; 91G20; 91G70; 91G60}% robustness paper: 62F35; 62F40; 91G20; 91G70
\newcommand{\jpJEL}{C52; C58; G12; C63; C12}% robustness paper:  C52; C58; C12; G12
\newcommand{\jpDateReceived}{26 July 2021} % {11 August 2021}
\newcommand{\jpDateRevised}{2 June 2023}

\newcommand{\jpDate}{}%\today

\author[1]{Jan Matas} % ORCID: ?
\author[1]{Jan Posp\'{\i}\v{s}il\thanks{Corresponding author, \email{honik@kma.zcu.cz}}} % ORCID: 0000-0002-4288-1614
\affil[1]{NTIS - New Technologies for the Information Society, Faculty of Applied Sciences, \authorcr University of West Bohemia, Univerzitn\'{\i} 2732/8, 301 00 Plze\v{n}, Czech Republic,\vspace*{3pt}}

\ifpdf
\hypersetup{
  pdftitle={\jpTitle},
  pdfauthor={\jpAuthors},
  pdfkeywords={\jpKeywords},
  pdfinfo={
      MSC={\jpMSC},
      JEL={\jpJEL}
  }
}
\fi

\title{\textcolor{Navy}{\textsc{\jpTitle}}}
\date{\jpDate}

\begin{document}

\maketitle
\vspace*{-3\baselineskip}
\begin{center}
\received{\jpDateReceived}
\revised{\jpDateRevised}
\end{center}

\begin{abstract}
% ------------------------------------------------------------------------------ begin: abstract.tex
In this paper, we analyze the robustness and sensitivity of various continuous-time rough Volterra stochastic volatility models in relation to the process of market calibration. Model robustness is examined from two perspectives: the sensitivity of option price estimates and the sensitivity of parameter estimates to changes in the option data structure. The following sensitivity analysis consists of statistical tests to determine whether a given studied model is sensitive to changes in the option data structure based on the distribution of parameter estimates. Empirical study is performed on a data set consisting of Apple Inc. equity options traded on four different days in April and May 2015. In particular, the results for RFSV, rBergomi and $\alpha$RFSV models are provided and compared to the results for Heston, Bates, and AFSVJD models.

% ------------------------------------------------------------------------------ end: abstract.tex
\end{abstract}

\ccode{MSC classification}{\jpMSC}
\ccode{JEL classification}{\jpJEL}

\setcounter{tocdepth}{2}
\tableofcontents

% ------------------------------------------------------------------------------ begin: introduction.tex
\section{Introduction}\label{sec:introduction}

In the field of mathematical finance, the stochastic volatility (SV) models are widely used to analyze derivative securities such as options. The SV models do not only assume that the asset price follows a specific stochastic process but also that the instantaneous volatility of asset returns is of random nature. 

The origin of these models goes back to the paper by \citet{HullWhite87} however the SV models became particularly popular thanks to the model by \citet{Heston93}, in which the volatility is modeled by the mean-reverting square root process. This model became popular among both practitioners and academics. 

Although many other SV models have been proposed since then, it seems that none of them can be considered to be the universally best market practice approach. Some models may perform well when calibrated to real market data with complex volatility surfaces but at the same time, they can suffer from over-fitting or they might not be robust to the changes in the option data structure as it is described by \cite{PospisilSobotkaZiegler19ee}. Moreover, a model with a good fit to an implied volatility surface might not be in-line with the observed properties of the corresponding realized volatility time series. 

One severe limitation of the classical SV models might be for example the independence of increments of the driving Brownian motion. This motivated \citet{ComteRenault98}, \cite{Comte12}, and independently for example \cite{Alos07} to consider the fractional Brownian motion (fBm) as the driving process since the fBm is a generalization of the Brownian motion which allows correlation of increments dependent on the so-called Hurst index $H \in (0,1)$. For $H>1/2$, the increments are positively correlated and we say that the process has long memory. For $H<1/2$, the increments are negatively correlated and we speak about short memory or more recently about "rough regime". \cite{Gatheral18} showed empirically that $H<1/2$ by estimating it from the realized volatility time series of major stock indexes and argues that the rough fractional stochastic volatility (RFSV) model is more consistent with the reality. 

In this paper, we consider the $\alpha$RFSV model recently introduced by \cite{MerinoPospisilSobotkaSottinenVives21ijtaf}. This model unifies and generalizes the RFSV model ($\alpha = 1$) and the rBergomi model ($\alpha = 0$). For pricing of a European call, we employ Monte-Carlo (MC) simulations using the Cholesky method equipped with the control variate variance reduction technique as it is suggested by \cite{MatasPospisil21roughsim}.

We then calibrate the model to a real market dataset and analyze its robustness to the changes in option data structure (options of different combinations of strikes and expiration dates may be available for trading on different days) using the methodology proposed by \cite{PospisilSobotkaZiegler19ee} which is based on data bootstrapping. In this paper, the authors showed that pricing using the classical SV models such as \cite{Heston93} and \cite{Bates96} models is highly sensitive to changes in option data structure. More robust results were obtained for the long-memory approximative fractional SV model, but not for all considered datasets. Then, a natural question arises: can the RFSV models perform better? Apparently, the answer is yes as we show in this paper. Since the RFSV models belong to a wider class of the rough Volterra processes, the presented methodology is applicable to this wider class as well. 

The structure of the paper is the following. In Section~\ref{sec:preliminaries}, we introduce the pertinent rough Volterra stochastic volatility models. 
In Section~\ref{sec:methods}, we describe the methodology, in particular the calibration of considered models to real market data. We describe the methodology of option data bootstrapping, as well as the details of the robustness and sensitivity analyses.
In Section~\ref{s:result:calib}, we summarize the obtained calibration results by comparing all the models in terms of variation in model parameters and in bootstrapped option model prices. We also test the roughness parameter and the parameter $\alpha$ for significance. Then, we provide the results of the sensitivity analysis fulfilled by a Monte Carlo filtering technique, testing whether a given studied model is sensitive to the changes in the option data structure when being calibrated.
We conclude all the obtained results in Section~\ref{sec:conclusion}.

% ------------------------------------------------------------------------------ end: introduction.tex
% ------------------------------------------------------------------------------ begin: preliminaries.tex
\section{Preliminaries and notation}\label{sec:preliminaries}

\subsection{Volterra volatility process}

Let $W=(W_t,t\geq 0)$ be a standard Wiener process defined on a probability space $(\Omega, \mathcal{F}, \Q)$ and let $\F^W=(\F^W_t,t\geq0)$ be the filtration generated by $W$. We consider a \emph{general Volterra volatility process} defined as
\begin{equation} \label{e:volprocess}
\sigma_t := g(t, Y_t), \quad t\geq 0,
\end{equation}
where $g: [0,+\infty) \times \mathbb{R} \mapsto [0,+\infty)$ is a deterministic function such that $\sigma_t$ belongs to $L^{1}(\Omega \times [0,+\infty))$ and $Y = (Y_t, t\geq 0)$ is the \emph{Gaussian Volterra process}
\begin{equation} \label{e:Volterra}
Y_t = \int_0^t K(t,s)\,\d W_s,
\end{equation}
where $K(t,s)$ is a kernel such that for all $t>0$
\begin{equation}
\int_0^t K^2(t,s) \d s < \infty, \tag{A1}\label{A1}
\end{equation}
and
\begin{equation}
\F^Y_t = \F^W_t. \tag{A2}\label{A2}
\end{equation}
By $r(t,s)$ we denote the autocovariance function of $Y_t$ and by $r(t)$ the variance
\begin{align}
r(t,s) &:= \E[Y_t Y_s], \quad t,s\geq 0, \notag \\
r(t) &:= r(t,t) = \E[Y_t^2], \quad t\geq 0. \label{e:r}
\end{align}

In particular we will model volatility as the \emph{exponential Volterra volatility process}
\begin{equation}\label{e:expVolterra}
\sigma_t = g(t,Y_t) = \sigma_{0}\exp\left\{\xi Y_t \NEW{- \frac12 \alpha\xi^2 r(t)} \right\}, \quad t\geq 0,
\end{equation}
where $(Y_t,t\geq 0)$ is the Gaussian Volterra process \eqref{e:Volterra} satisfying assumptions \eqref{A1} and \eqref{A2}, $r(t)$ is its autocovariance function \eqref{e:r}, and $\sigma_0>0$, $\xi>0$ and $\alpha\in[0,1]$ are model parameters.

A very important example of Gaussian Volterra processes is the \emph{standard fractional Brownian motion} (fBm) $B^{H}_{t}$ (the exponent $H$ has the meaning of index, not power)
\begin{equation} \label{e:fBm}
B_t^H = \int_0^t K(t,s)\,\d W_s,
\end{equation}
where $K(t,s)$ is a kernel that depends also on the Hurst parameter $H \in (0,1)$. Recall that the autocovariance function of $B^{H}_{t}$ is given by
\begin{equation}\label{e:fBm_r}
r(t,s):=\E[B^H_t B^H_s] = \frac12 \left( t^{2H} + s^{2H} - |t-s|^{2H}\right), \quad t,s\geq 0,
\end{equation}
and in particular $r(t):=r(t,t) = t^{2H}$, $t\geq0$.

Nowadays, the most precise Volterra representation of fBm is the one by \cite{Molchan69}
\begin{align}
B^H_t &:= \int_0^t K_H(t,s) \d W_s, \label{e:Molchan_fBm}
\intertext{where}
K_H(t,s) &:= C_H \Biggl[ \left( \frac{t}{s} \right)^{H-\frac12} (t-s)^{H-\frac12} %\notag \\
- \left(H-\frac12\right) s^{H-\frac12} \int_s^t z^{H-\frac32} (z-s)^{H-\frac12} \d z \Biggr] \label{e:Molchan_kernel} \\
C_H &:= \sqrt{\frac{2 H \Gamma\left(\frac32-H\right)}{\Gamma\left(H+\frac12\right)\Gamma\left(2-2H\right)} }. \notag
\end{align}
To understand the connection between Molchan-Golosov and other representations of fBm such as the original \cite{Mandelbrot68} representation, we refer readers to the paper by \cite{Jost08}. 

There are various methods to simulate the fractional Brownian motion numerically. We often divide these methods into two classes: exact methods and approximate methods \citep{Dieker02}. We focus on more accurate exact methods that usually exploit the covariance function \eqref{e:fBm_r} of the fBm to simulate exactly the fBm (the output of the method is a sampled realization of the fBm) without the necessity to treat the complicated Volterra kernel. In particular, the Cholesky method use a covariance matrix to generate the fBm from two independent normal samples. Despite its higher computational complexity, this method has already proved to be the most suitable for simulation of the volatility models \citep{MatasPospisil21roughsim}.

\subsection{Rough Volterra volatility models}

Let $S=(S_{t}, t\in[0,T])$ be a strictly positive asset price process under a market chosen risk neutral probability measure $\Q $ that follows the stochastic dynamics:

\begin{eqnarray}\label{e:model}
\d S_{t}= r S_{t} \d t + \sigma_{t}S_{t}  \left(\rho \d W_{t}  + \sqrt{1-\rho^{2}}\d \widetilde{W}_{t} \right),
\end{eqnarray}
where $S_{0}$ is the current \emph{spot} price, $r\geq 0$ is the all-in interest rate, $W_{t}$ and $\widetilde{W}_{t}$ are independent standard Wiener processes defined on a probability space $(\Omega, \mathcal{F}, \Q)$ and $\rho\in[-1,1]$ represents the correlation between $W_t$ and $\widetilde{W}_{t}$.

Let $\mathcal{F}^{W}$ and $\mathcal{F}^{\widetilde{W}}$ be the filtrations generated by $W$ and $\widetilde{W}$ respectively and let $\mathcal{F}:=\mathcal{F}^{W} \vee \mathcal{F}^{\widetilde{W}}$. The \emph{stochastic volatility process} $\sigma_{t}$ is a square-integrable Volterra process assumed to be adapted to the filtration generated by $W$ and its trajectories are assumed to be a.s. c\`adl\`ag and strictly positive a.e. exponential Volterra volatility process satisfies these properties). 

For convenience we let $X_{t} = \ln S_{t}$, $t\in[0,T]$, and consider the model
\begin{eqnarray}\label{e:log-model}
\d X_{t}= \left(r -\frac{1}{2}\sigma^{2}_{t}\right)\d t + \sigma_{t} \left(\rho \d W_{t}  + \sqrt{1-\rho^{2}} \d \widetilde{W}_{t} \right).
\end{eqnarray}
Recall that $Z:= \rho  W  + \sqrt{1-\rho^{2}} \widetilde{W}$ is a standard Wiener process.

In this paper we will study the {$\alpha$RFSV} model firstly introduced by \cite{MerinoPospisilSobotkaSottinenVives21ijtaf}. In this model, the volatility is modelled as the exponential Volterra process with fBm, i.e. 
\begin{equation}\label{e:sigma_fBm}
\sigma_t = \sigma_0\exp\left\{\xi B^H_t \NEW{- \frac12\alpha\xi^2 r(t)} \right\}, \quad t\geq 0,
\end{equation}
where $\sigma_0>0$, $\xi>0$ and $\alpha\in[0,1]$ are model parameters together with $H<1/2$ that guarantees the \emph{rough} regime. For $\alpha = 0$ we get the {RFSV model} \citep{Gatheral18}, for $\alpha = 1$ the {rBergomi} model \citep{Bayer16}.

While both cases of the $\alpha$RFSV model are more likely to replicate the stylized facts of volatility \citep{Gatheral18} even by using relatively small number of parameters ($\sigma_0, \xi, \rho, H$), the issue is the non-markovianity of the model. Because of this, we cannot derive any semi-closed form solution using the standard It\^o calculus nor the Heston's framework. Therefore, to price even vanilla options, we have to rely on Monte-Carlo (MC) simulations. For these purposes, a modified Cholesky method will be used together with the control variate variance reduction technique as it was described by \cite{MatasPospisil21roughsim}.

We close this section by mentioning, that there exists yet another pricing approach that takes advantages of the so-called approximation formula derived by \cite{MerinoPospisilSobotkaSottinenVives21ijtaf}. This formula can be used either as a standalone fast approximation or together with the MC simulations to speed up the calibration tasks. However, in this paper, we will focus on robustness and sensitivity analyses based on pricing approaches that are as accurate as possible and this can be achieved currently only by the MC simulations that use an exact simulation technique for the fBm.

% ------------------------------------------------------------------------------ end: preliminaries.tex
% ------------------------------------------------------------------------------ begin: methods.tex
\section{Methodology}\label{sec:methods}

In this section, we describe the methodology of calibration of the rough Volterra models to real market data and we focus on the robustness and sensitivity analysis.

\subsection{Calibration to market data}\label{s:meth:calib}

Model calibration constitutes a way to estimate model parameters from available market data. The alternative approach suggests estimating the parameters directly from time series data such as for example \cite{Gatheral18} did for the Hurst parameter. We understand model calibration as the problem of estimating the model parameters by fitting the model to market data with pre-agreed accuracy.

Mathematically, we express the calibration problem as an optimization problem
\begin{align}\label{e:calibration}
\inf_\Theta G(\Theta), \quad G(\Theta) = \sum_{i=1}^N w_i[C_i^\Theta(T_i, K_i) - C_i^\text{mkt}(T_i, K_i) ]^2,
\end{align}
where $C_i^\text{mkt}(T_i, K_i)$ is the observed market price of the $i$th option, $i=1,\ldots,N$, with time to maturity $T_i$ and of strike price $K_i$, while $w_i$ is a weight and $C^\Theta(T_i, K_i)$ denotes the option price computed under the model with a vector of parameters $\Theta$. For $\alpha$RFSV, we have $\Theta = [\sigma_0, \rho, H, \xi, \alpha]$.

In fact, the representation of the calibration problem in \eqref{e:calibration} is a non-linear weighted least squares problem. To obtain a reasonable output, we have to assume that the market prices are correct, i.e., there is no inefficiency in the prices, which is usually not the case, especially for options being further ITM or OTM. To fix this, let us assume that the more an option is traded, the more accurate the price is. We can then weight the importance of a given option in the least squares problem by the traded volume of the given option. However, there is also another, and in fact a more convenient and popular way to implement such weights. We can get the information of uncertainty about the price of an option from its bid-ask spread. %Bid price is the amount of money a buyer is willing to pay for buying the option and otherwise, ask price is the one, a buyer is willing to sell at. 
The greater the bid-ask spread, the more uncertainty (and usually less trading volume) there is about the price. Therefore, we will use a function of bid and ask prices for the weights: $w_i = g(C_i^{\text{bid}} - C_i^{\text{ask}})$, where $g(x)$ can be, for example, $1/{x^2}, 1/\abs{x}, 1/\sqrt{x}$, etc. Based on the empirical results \citep{MrazekPospisilSobotka16ejor}, we will consider only the case $g(x)=1/x^2$.  

Because the objective function is non-linear, we cannot solve the problem analytically as in the case of standard linear regression. Hence, we revert to iterative numerical optimizers.

For the minimization of \eqref{e:calibration}, we use the MATLAB function \texttt{lsqnonlin()} that implements an interior trust region algorithm, described by \cite{Coleman96}. The algorithm assumes, among other things, that the target function is convex. However, we cannot even show the convexity of the target function since we have no analytical expression to describe it. Therefore, if the algorithm ends up in a local minimum, it is not guaranteed that it is the global minimum. 

In fact, the target function can have more than one local minimum (the source is the non-linearity of the model price function). To determine the initial point for gradient-based \texttt{lsqnonlin()}, we use another MATLAB function \texttt{ga()} that implements a genetic algorithm minimization approach. It deploys a predefined number\footnote{We use 150 points.} of initial points across the domain of the function and then, each point serves as an initial condition for minimization that is performed for a pre-defined number of steps. Based on the genetic rules of random mutation, crossbreeding, and preservation of the fittest, the most successful points are preserved, perturbated by a random mutation, and crossbred among themselves. This approach \citep{MrazekPospisilSobotka16ejor} has been shown to produce sound results.

To measure the quality of the fit of a calibrated model, we use the following metrics. Having $N$ options in the data set, we denote $C^\text{mkt}_i$ the market price of the $i$th option and $\tilde{C}_i$ the estimated price of the $i$th option based on the calibrated model. Denoting $S_0$ the spot price, the first metric is the \emph{average relative fair value} (ARFV) and the second one is the \emph{maximum relative fair value} (MRFV). They can be expressed as
\begin{equation*}
ARFV =  \frac{1}{N} \sum_{i=1}^N \frac{\abs{  \tilde{C}_i  - C^\text{mkt}_i }}{ S_0}, \qquad
MRFV =  \max_{i=1,\ldots,N} \frac{\abs{  \tilde{C}_i  - C^\text{mkt}_i }}{ S_0}.
\end{equation*}
It is worth to mention that these measures offer a better error understanding than the originally used \emph{average absolute relative error} (AARE) and \emph{maximum absolute relative error} (MARE)
\begin{equation*}
AARE = \frac{1}{N} \sum_{i=1}^N \frac{\abs{  \tilde{C}_i  - C^\text{mkt}_i }}{C^\text{mkt}_i}, \qquad
MARE = \max_{i=1,\ldots,N} \frac{\abs{  \tilde{C}_i  - C^\text{mkt}_i }}{C^\text{mkt}_i}.
\end{equation*}

\subsection{Robustness analysis} \label{s:meth:calib:robAn}

We calibrate the $\alpha$RFSV model the way described in the previous section to a real market dataset. In the ideal hypothetical case, all combinations of strikes and times to maturity for a given option would be available, i.e., we would have a continuous price surface to which we would calibrate a selected model. However, in reality, we have only a finite number of different options available to trade and moreover, the combinations of strikes and times to maturities (we call this option data structure) changes and even the number of combinations itself changes over time. Therefore, the obtained coefficient estimates can differ, should the model calibration be sensitive to the option data structure. 

In this paper, we understand \emph{robustness} as the property of a model that conveys the sensitivity of the model being calibrated to changes in the option structure. To study the robustness of the $\alpha$RFSV model, we use the methodology suggested by \cite{PospisilSobotkaZiegler19ee}. Therefore, our results of the robustness analysis of the $\alpha$RFSV model are comparable with those of the Heston, Bates, and the approximate fractional stochastic volatility jump diffusion (AFSVJD) model, presented in the referenced paper\footnote{Not to be confused, please note that a shorter abbreviation FSV is used therein.}.

To analyze robustness, we have to simulate the changes in the option structure. To do this, we employ bootstrapping of a given option structure. Bootstrapping is a technique when random samples are selected with replacement from the initial dataset. For example, to bootstrap the data set $(X_1, X_2, \ldots, X_6)$, we need to generate uniformly distributed random integers from $\{1,2,\ldots,6\}$. Suppose the realization is $\{2,3,5,4,4,3\}$. Then, the obtained bootstrapped sample is $(X_2, X_1, X_5, X_4, X_4, X_3)$. 

Mathematically, an option structure is the set of all the combinations of strikes $K$ and times to maturity $T$ available for trading in a given day. Having market data consisting of $N$ options, the set $\mathcal{X} = \{(K_i, T_i), i = 1, \ldots, N \}$ is the option structure for the given day where each option has the market price $C^\text{mkt}_i = C^\text{mkt}(K_i, T_i)$. 

By bootstrapping $\mathcal{X}$ in total of $M$ times, we obtain $M$ new option structures $\mathcal{X}_1, \ldots, \mathcal{X}_M$. Then each $\mathcal{X}_j$, together with the option prices from the initial dataset assigned to the corresponding combinations of strikes and times to maturities, produces bootstrapped sample $B_j$. Next, we calibrate the model separately to each $B_j$ and obtain estimates of the model parameters and model prices. Let us denote $\tilde{\Theta}_j$ the parameter estimates obtained from the bootstrapped sample $B_j$, and $\tilde{C}^j=[\tilde{C}^j_1, \ldots, \tilde{C}^j_N]$, where $\tilde{C}^j_i = \tilde{C}^j_i(K_i,T_i)$, is the vector of corresponding model prices. 

Having the results of the calibrations from $B_1, \ldots, B_M$, we can compute the bootstrap estimates of the parameters and models prices. The bootstrap estimate of a parameter is the mean across all the estimated parameters: 
\begin{equation} \label{e:meth:calib:boot_mean_coeff}
\hat{\Theta} = \frac{1}{M}\sum_{i=1}^{M}\tilde{\Theta}_i
\end{equation}
and the bootstrap estimate of a model price of the $i$th option is
\begin{equation*}
\hat{C}_i = \frac{1}{M}\sum_{j=1}^{M} \tilde{C}^j_i.
\end{equation*}
Next, we look at the variance of the errors of the price estimates of the $i$th option $\abs{\tilde{C}^j_i - C^\text{mkt}_i  }$. However, to be able to better compare the variances among different options, we normalize the error. Then, let us denote 
\begin{equation} \label{e:meth:calib:robAn:Variance_V_i}
V_i = \Var \left[  \frac{\abs{\tilde{C}^j_i - C^\text{mkt}_i  }}{C^\text{mkt}_i}   \right]
\end{equation}
the variance of the normalized errors of the $i$th option.
It is also useful to examine the \emph{bootstrap relative error} (BRE) for the $i$th option:
\begin{equation} \label{e:meth:calib:robAn:BRE}
BRE_i = \frac{\abs{\hat{C}_i - C^\text{mkt}_i } }{   C^\text{mkt}_i  } = \frac{\abs{\frac{1}{M}\sum_{j=1}^{M} ( \tilde{C}^j_i - C^\text{mkt}_i )} }{   C^\text{mkt}_i  } .
\end{equation}

We analyze variation in coefficients visually by plotting a scatter plot matrices. Denoting $d$ the number of model coefficients being calibrated, the scatter plot matrix is a $d \times d$ matrix, where histograms for each coefficient are on the diagonal, and 2D scatter plots of corresponding values of coefficients elsewhere. Hence, from a scatter plot matrix, we get a grasp of the distributions of coefficients and also whether there is any dependence between pairs of coefficients and variation in the estimates.

\subsection{Sensitivity analysis} \label{s:meth:calib:sensAn}

In this paper, we use a similar method to carry out a sensitivity analysis introduced by \cite{PospisilSobotkaZiegler19ee} based on the ideas of \cite{Saltelli08}. In short, we aim to test whether the $\alpha$RFSV model is sensitive to changes in option structure through a given parameter. 

In our context, we chose the following Monte-Carlo filtering technique\footnote{For more details on Monte-Carlo filtering approaches see, for instance \cite{Saltelli08}.}: To each vector of calibrated model parameters obtained from the bootstrapped data, we calculate the average relative fair value (ARFV) as a quality measure for the calibrated model fit. Then, we separate the calibrated models into three groups: (I) the calibrated models with the corresponding values of the ARFV up to the third octile, (II) the models with the ARFV between the third and the fifth octile, and (III) the models with the ARFV above the fifth octile. Next, for each calibrated parameter, we compare the distribution of the parameter estimates corresponding to models from group (I) with the distribution from group (III). We use the Kolmogorov-Smirnov test for the comparison. The null hypothesis is that the parameter estimates from group (I) comes from the same distribution as those from group (III).

% ------------------------------------------------------------------------------ end: methods.tex
% ------------------------------------------------------------------------------ begin: results.tex
\section{Numerical results} \label{s:result:calib}

In this section, we present our results of calibrations and robustness and sensitivity analyses of the $\alpha$RFSV model. We used the same real market dataset as in the paper by \cite{PospisilSobotkaZiegler19ee} where Heston, Bates, and the AFSVJD models were analyzed. Consequently, the results are directly comparable.

\subsection{Data description} \label{s:res:calib:data}

We use a real market dataset that consists of market prices of call options on Apple Inc. stock (NASDAQ: AAPL) quoted on four days of 2015: 04/01, 04/15, 05/01, 05/15. Naturally, the combinations of strikes and times to maturity of the options (the option data structure) change over time. There are 113 options in the option chain on the first day. The second day, the total number of different options rises to 158, the next day to 201, and the last day decreases to 194. %This is a common phenomenon caused by issuers of derivatives reacting on moves in the spot price of a stock or on new information being revealed.

\begin{figure}[hb!]
\centering
\includegraphics[width=12cm]{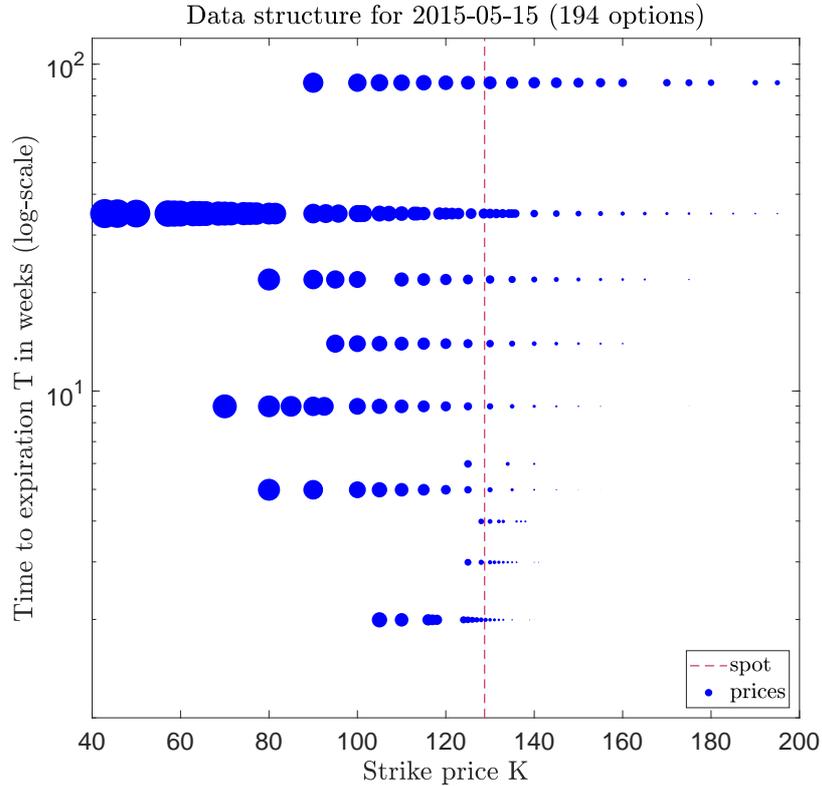}
\caption{Call option data structure for AAPL dataset from May, 15, 2015. The positions of disks are given by the combinations of the strikes $K$ on the $x$-axis and the maturities $T$ on the $y$-axis of the options listed at the time. The diameter of each disk relates to the corresponding close price. 
\label{f:results:data:option_strucure}
}
\end{figure}

For convenience we visualize the data from May, 15, in Figure \ref{f:results:data:option_strucure}, in order to give some perspective. For each listed call option with the strike $K$ and time to maturity $T$, a disk is plotted with center in $(K,T)$. The diameter of the disk relates to the price of the option.

\subsection{Calibration routine} \label{s:result:calib:routine}

In order to calibrate the $\alpha$RFSV model, we use the Cholesky method with the modified turbocharging method introduced by \cite{MatasPospisil21roughsim} and we follow the recommendation given there to employ at least $P=150,000$ paths discretized by $n = 4\times252$ per interval $[0,1]$, so the pricing method assures sufficient accuracy. Then, we are able to price any option with $T < T_{\text{max}}$ from the paths, which were already simulated, by truncating them to corresponding interval $[0, T]$.

\begin{figure}[ht!]
\centering
\includegraphics[width=12cm]{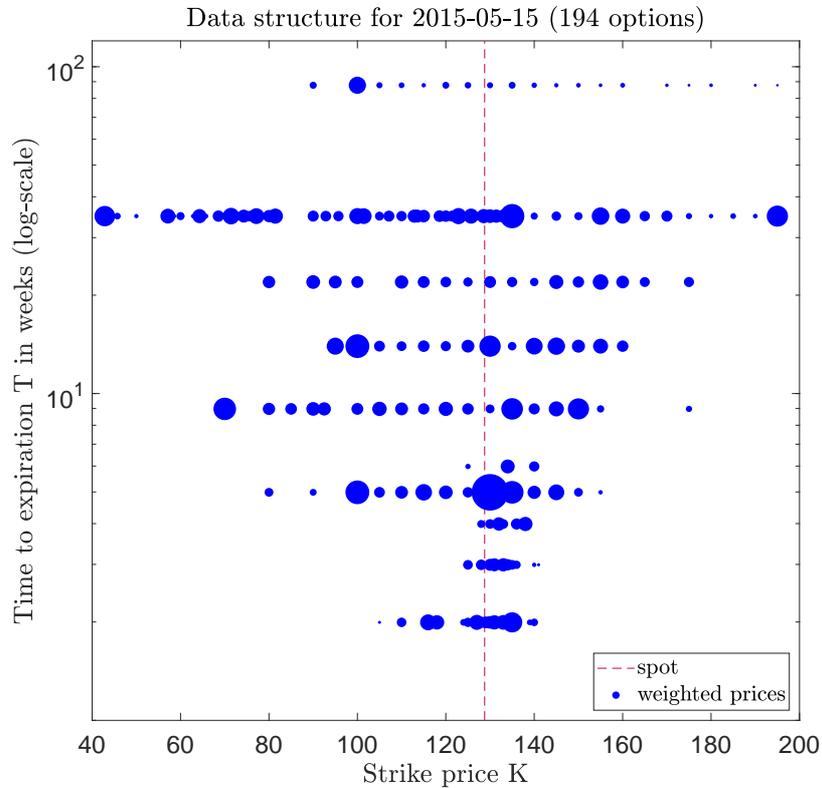}
\caption{Example of the options call data structure for AAPL call option prices from May, 15, 2015, weighted by \eqref{e:results:calib:routine:weight}. The positions of disks are given by the combinations of the strikes $K$ on the $x$-axis and the maturities $T$ on the $y$-axis of the options listed at the time. The diameter of each disk relates to the corresponding weighted close price. Compare with Figure \ref{f:results:data:option_strucure}. 
\label{f:results:data:option_strucure:weightedPrices}
}
\end{figure}

We tried to use different weights\footnote{Having $w_i = g(C_i^{\text{bid}} - C_i^{\text{ask}})$, we tried $g(x) = \frac{1}{x^2}, g(x) = \frac{1}{\abs{x}}$, and $ g(x) = \frac{1}{\sqrt{x}}$.} for the target function \eqref{e:calibration}, but the best results were obtained for the weight type
\begin{equation}\label{e:results:calib:routine:weight}
w_i = \frac{1}{\left( C_i^\text{bid} - C_i^\text{ask}  \right)^2 },
\end{equation}
which aligns with the results by \cite{MrazekPospisilSobotka16ejor} for other SV models. For that reason, we present only the results for this type of weights. To compare weighted prices with the market prices, see Figure \ref{f:results:data:option_strucure:weightedPrices} and \ref{f:results:data:option_strucure}.

\subsection{Overall calibration} \label{s:result:calib:overall}

First, we summarize the results of the calibrations to the market data. Then, we compare the results to those obtained by \cite{PospisilSobotkaZiegler19ee} where different SV models were analyzed using similar methods and the identical dataset. For convenience of the comparison, we adopt Table 1 from the referenced paper as Table \ref{t:Overallerror}. It contains AAREs of the calibrated Heston, Bates, and AFSVJD models. Lastly, we test the significance of the $H$ and $\alpha$ parameters.
 
\begin{table}[ht!]
\centering
\caption{Average  ($AARE$) of overall calibrations of the Heston, Bates, and AFSVJD models for the same dataset as we use, reprint of \cite[Table 1]{PospisilSobotkaZiegler19ee}.}\label{t:Overallerror}
\begin{tabular}{rllll} %\toprule & \multicolumn{2}{c}{Initial settings for calibration} \\
Trading day & 1/4/2015 & 15/4/2015 & 1/5/2015 & 15/5/2015 \\
\midrule 
Heston & $5.15\%$ & $3.79\%$ & $6.58\%$ & $3.39\%$\\
Bates & $3.73\%$ & $3.57\%$ & $5.77\%$ & $3.41\%$\\
AFSVJD & $2.21\%$ & $2.16\%$ & $5.89\%$ & $3.20\%$\\
\bottomrule
\end{tabular}
\end{table}

For the MATLAB function \texttt{ga()}, we set the number of initial points on 150 and the number of iterations on 5, as more than 5 did not make much significant improvement. For \texttt{lsqnonlin()}, which is ran after \texttt{ga()} and which further minimizes its output, we set the tolerance on the value of the target function on $10^{-6}$ and the tolerance on the norm of the difference between two subsequent points on $10^{-7}$. Although the global optimization part is heavily time consuming, it is crucial in the situations when no initial guess is available to be used for the local optimization part that is significantly faster for obvious reasons. The whole procedure takes just a couple of minutes on a personal computer and no supercomputing power is necessary.

\begin{table}[hb!]
\centering
\caption{Lower and upper bounds for the model coefficients we considered for the overall calibration.}\label{t:res:calib:bound}
\begin{tabular}{@{}rlllll@{}}
\toprule
Coefficient     & $\sigma_0$ & $\rho$ & $H$   & $\xi$    & $\alpha$ \\ \midrule
Lower bound     & 0.01       & -1     & 0.05  & 0.01      & 0 \\
Upper bound     & 0.20       & -0.05  & 0.25  & 3        & 1 \\ \bottomrule
\end{tabular}
\end{table} 

The bounds of the coefficients considered for the calibration are summarized in Table \ref{t:res:calib:bound}. While the bounds $\sigma_0 > 0$ and $\rho > -1$ are naturally arising from the definition of the $\alpha$RFSV model, the upper bound for $\sigma_0$ and the bounds for $\xi$ were determined based on several test calibrations such that they provided a suitable area for the genetic algorithm while not limiting the calibration procedure in finding the global maximum. The upper bound $\rho \leq -0.05$ is based on the recommendation given in \cite{MatasPospisil21roughsim} and the range for the Hurst parameter was set as $0.05\leq H \leq 0.25$ which is, according to \cite{Bennedsen21}, a common range for $H$ based on estimates for 2000 different equities.

\begin{table}[ht!]
\centering
\caption{Average ($AARE$) of overall calibrations of the RFSV, rBergomi, and $\alpha$RFSV models.
\label{t:Overallerror_RFSV}
}
\begin{tabular}{rllll} %\toprule & \multicolumn{2}{c}{Initial settings for calibration} \\
Trading day & 1/4/2015 & 15/4/2015 & 1/5/2015 & 15/5/2015 \\
\midrule 
RFSV & $27.03\%$ & $7.00\%$ & $7.38\%$ & $7.48\%$\\
rBergomi & $6.46\%$ & $6.99\%$ & $9.74\%$ & $15.63\%$\\
$\alpha$RFSV & $5.74\%$ & $6.70\%$ & $9.71\%$ & $11.20\%$\\
\bottomrule
\end{tabular}
\end{table}

Table \ref{t:Overallerror_RFSV} presents the results of the overall calibration procedure for the studied models. We can see that for the first two days, rBergomi provides better fits than RFSV. For the next two days, the situation reverses and the fits obtained by RSFV are superior to those obtained by rBergomi. An interesting result is that the $\alpha$RFSV model that unifies the two models by introducing a new parameter $\alpha$ that fulfills the role of the weight between the models fits the data in the most consistent way. We discovered that for the first two days, the parameter $\alpha$ is closer to $1$ which corresponds to the rBergomi model and the next two days, $\alpha$ leans towards $0$ thus the RFSV model. However, the $\alpha$RFSV model does not always provide the best fit.

Comparing the values of AARE in Table \ref{t:Overallerror_RFSV} to those obtained for other SV models tabulated in Table \ref{t:Overallerror}, we can observe that the rough models do not provide better fits than the SV models. Nevertheless, rough models, as we will see in the next sections, are much more robust compared to the SV models.

\begin{figure}[ht!]
\centering
\includegraphics[width=.49\textwidth]{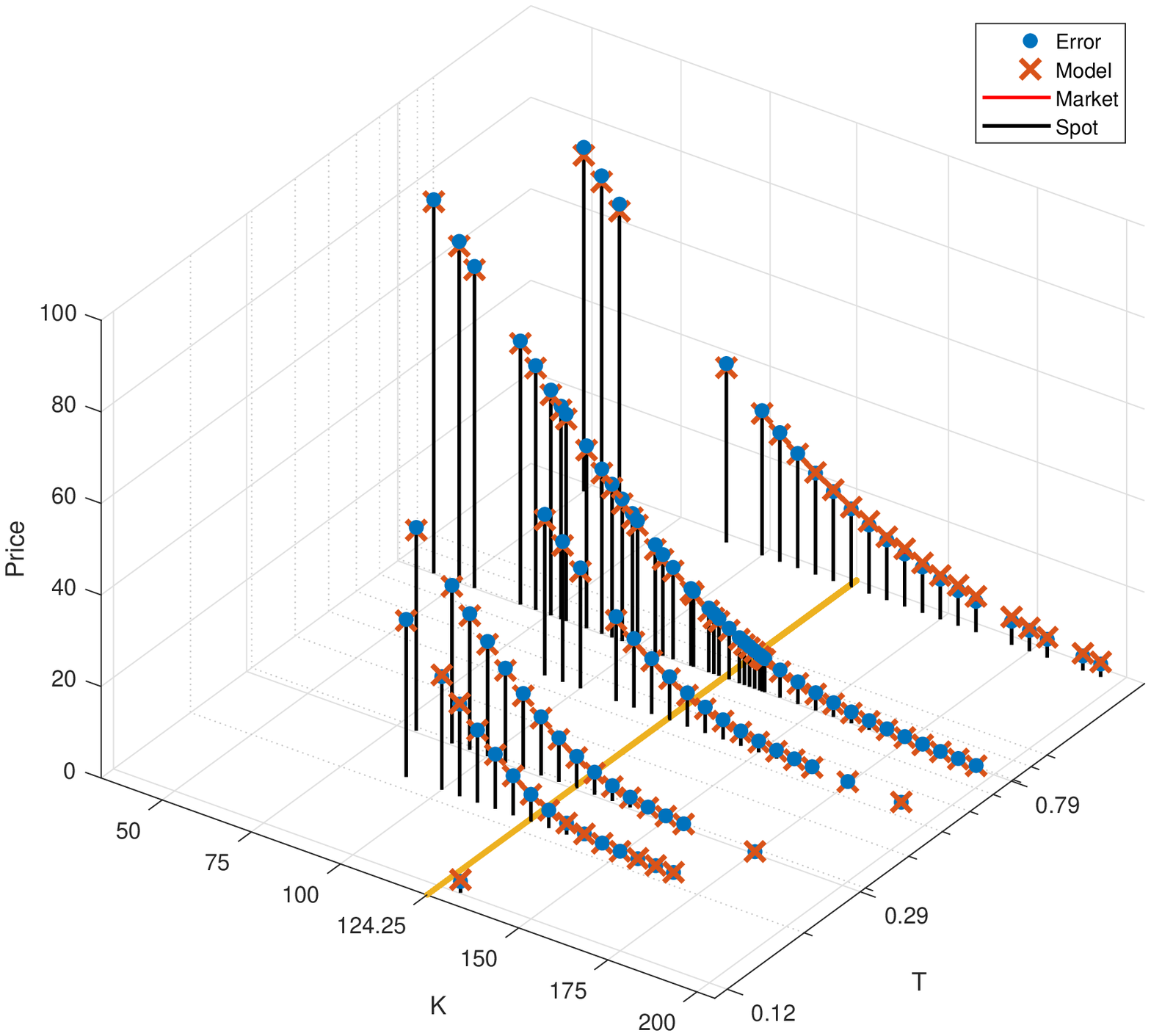}
\includegraphics[width=.49\textwidth]{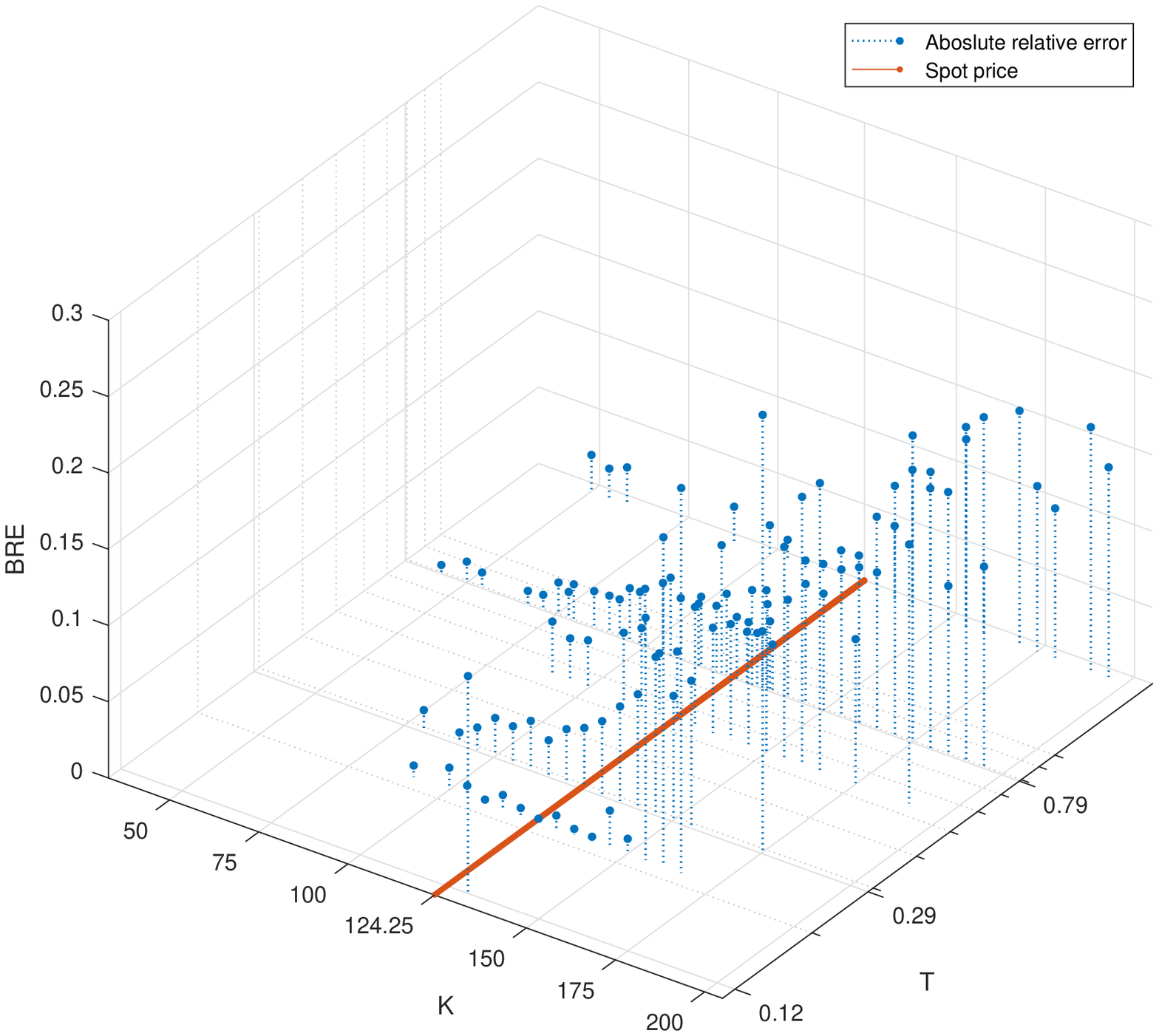}\\
\includegraphics[width=.49\textwidth]{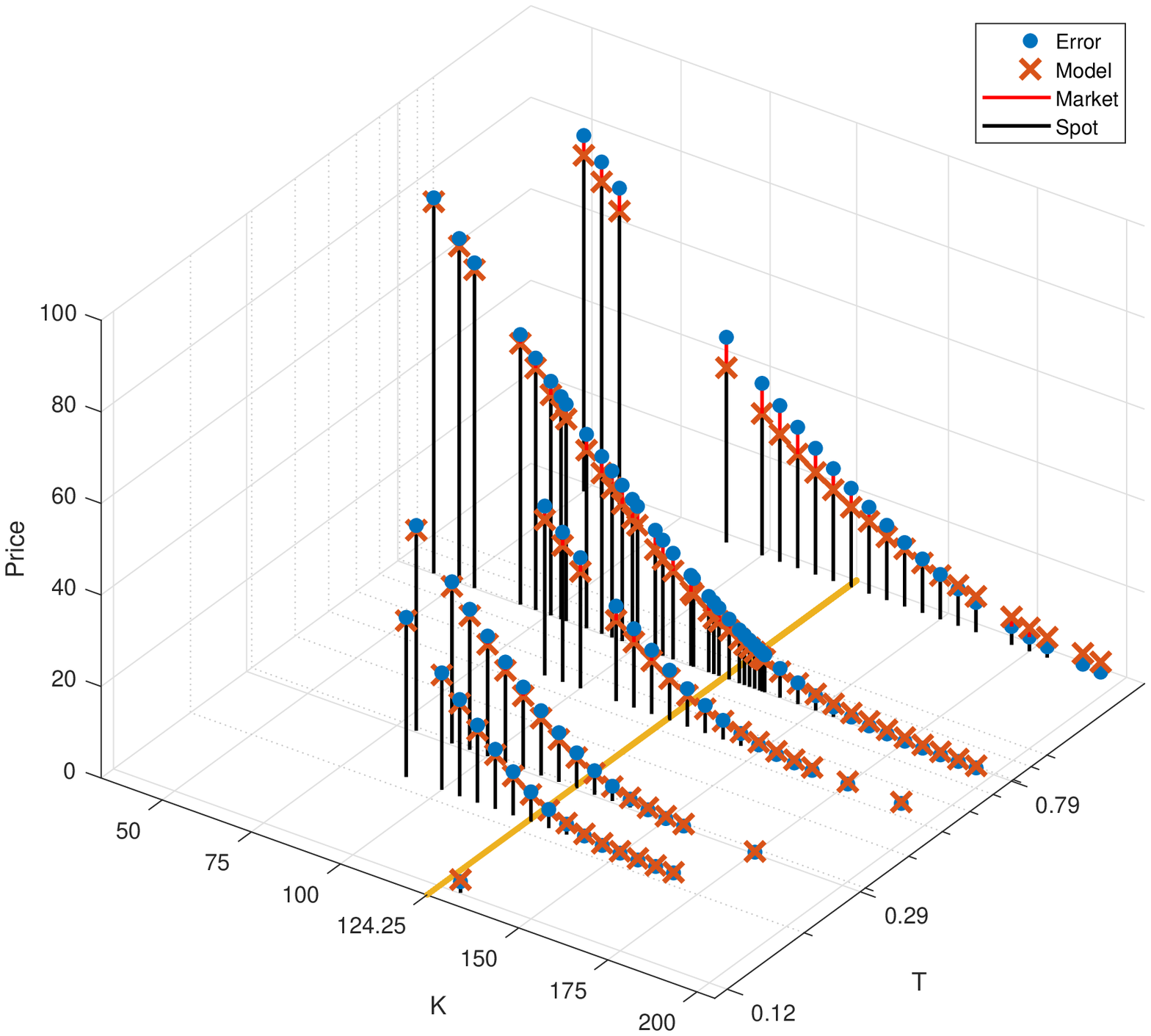}
\includegraphics[width=.49\textwidth]{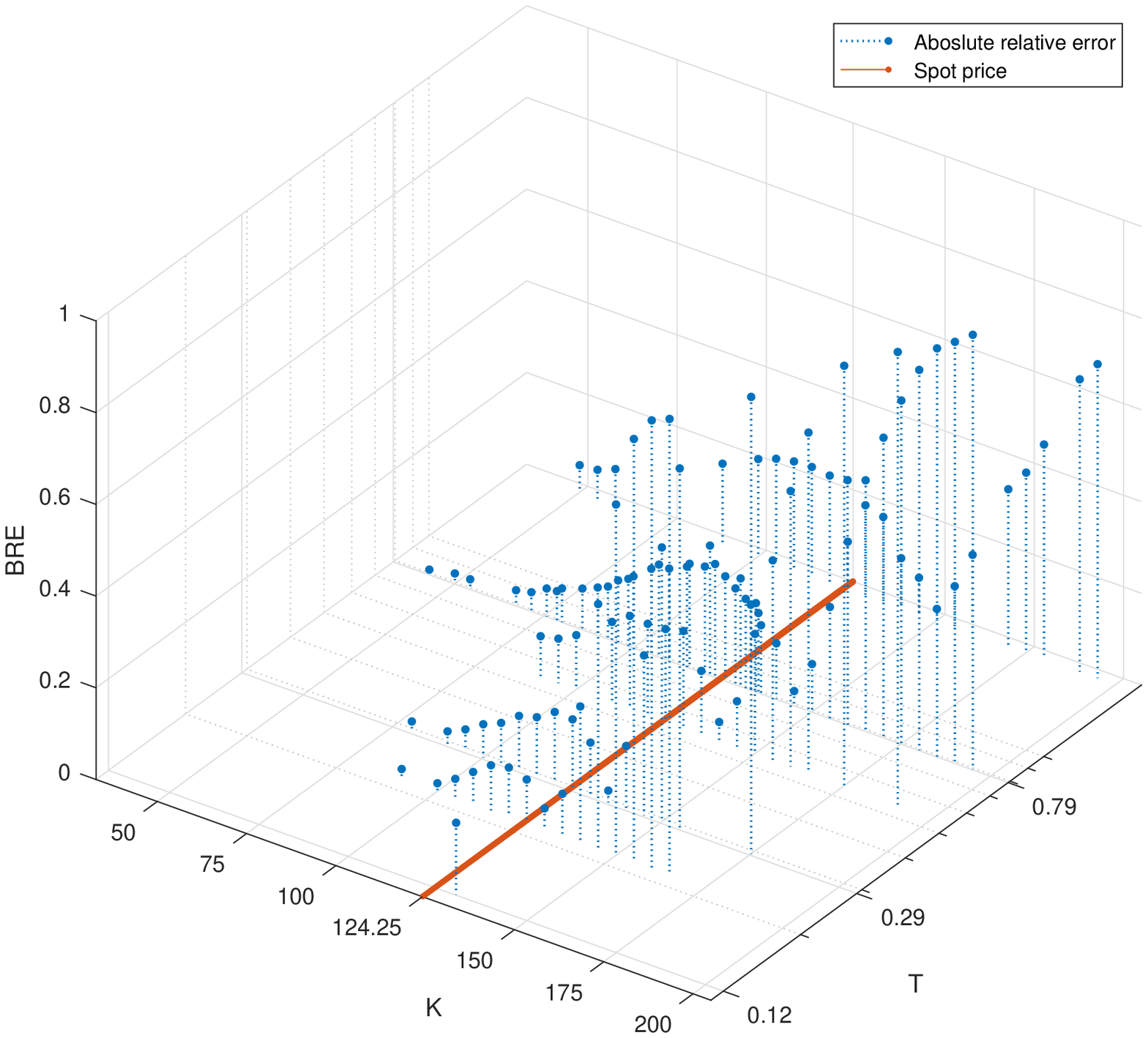}
\caption{The market and model prices (left) and the corresponding BREs (right) for the rBergomi model (top row) and the RFSV model (bottom row) on April 01. \label{f:overall:plot3D}}
\end{figure}

To illustrate the difference in the fit of two different models on one day, we visualized the model prices, the market prices, and the BREs in Figure \ref{f:overall:plot3D}. We chose April 01 because there is the biggest difference in the AARE of rBergomi (top row) and RFSV (bottom row). Above the $K-T$ plane we plotted the market and model prices together (left side) and the corresponding BREs (right side). On this day, the rBergomi model provides much better fit than the RFSV model.

\begin{table}
\centering
\caption{The overall calibration results.}\label{t:overall:results}
\begin{tabular}{llllllrrrr}
\hline
\multicolumn{10}{c}{Overall calibration of the RFSV model} \\ 
\hline
day & $\sigma_0$ & $\rho$ & $H$ & $\xi$ & $\alpha$ & $AARE$ & $MARE$ & $WRSS$ & $ARFV$ \\ 
\hline
4-01 &  0.0800 &  -0.9000 &  0.3000 &  1.5000 &  0 &    27.03\% &  99.65\% &  0.3381 & -- \\ 
4-15 &  0.0700 &  -0.2006 &  0.2258 &  0.0602 &  0 &     7.00\% &  68.78\% &  0.0430 & -- \\ 
5-01 &  0.0467 &  -0.2259 &  0.0740 &  0.9101 &  0 &     7.38\% &  79.41\% &  0.0492 & -- \\ 
5-15 &  0.0355 &  -0.2533 &  0.1748 &  1.3315 &  0 &     7.48\% &  53.78\% &  0.0123 & -- \\ 
\hline
\multicolumn{10}{c}{Overall calibration of the rBergomi model} \\ 
\hline
4-01 &  0.0782 &  -0.1792 &  0.2324 &  0.9875 &  1 &     6.46\% &  28.60\% &  0.0226 & 0.3983\%\\ 
4-15 &  0.0700 &  -0.1771 &  0.0518 &  0.8858 &  1 &     6.99\% &  48.92\% &  0.0282 & 0.2825\% \\ 
5-01 &  0.0615 &  -0.0755 &  0.1047 &  0.3520 &  1 &     9.74\% &  94.17\% &  0.0516 & 0.4439\% \\ 
5-15 &  0.0470 &  -0.1243 &  0.0634 &  0.3126 &  1 &     15.63\% &  107.46\% &  0.0443 & 0.4111\% \\ 
\hline
\multicolumn{10}{c}{Overall calibration of the $\alpha$RFSV model} \\ 
\hline
4-01 &  0.0714 &  -0.1830 &  0.2336 &  0.8229 &  0.8213 &  5.74\% &  49.76\% &  0.0286 & 0.3357\% \\ 
4-15 &  0.0714 &  -0.1830 &  0.1434 &  0.3910 &  0.9721 &  6.70\% &  64.83\% &  0.0384 & 0.2683\% \\ 
5-01 &  0.0553 &  -0.0578 &  0.1038 &  0.9510 &  0.3796 &  9.71\% &  55.56\% &  0.0522  & 0.4451\% \\ 
5-15 &  0.0433 &  -0.3302 &  0.1607 &  1.1077 &  0.2585 &  11.20\% &  80.82\% &  0.0247  & 0.2800\% \\ 
\end{tabular}
\end{table}

\subsubsection{Parameter significance testing}
We also test for parameter significance. We are particularly interested whether the parameters $H$ and $\alpha$ have any affect on the model fit, i.e., whether the fit of the $\alpha$RFSV model is better/worse when $H$ (resp. $\alpha$) is being calibrated compared to the model with fixed $H$ (resp. $\alpha$). We consider the fixed value of $H$ being $1/2$, thus it constituted a model with the volatility process being driven just by the Bm instead of the fBm. To test the significance of $\alpha$, we compare the fit of the $\alpha$RFSV model with the rBergomi model which corresponds to $\alpha = 1$.

If we had a deterministic pricing formula, we could simply calibrate the models and compare the fit directly. But since the pricing involves randomness (Monte Carlo simulations), we need to conduct a statistical test to decide whether the difference in the fit measured by the ARFV is significant. We thus conducted 100 simulations to that resulted in different prices and thus a sample of different values of the ARFV for a given calibrated model. We then used the two-sample t-test to compare the mentioned pairs of models. 

We first used this method to compare the rBergomi model with $H$ being calibrate and with $H$ fixed to $1/2$. For all the four days, rBergomi provided significantly better fit. Then we compared the $\alpha$RFSV model with the rBergomi model which corresponds to $\alpha = 1$. Again, the null hypothesis was rejected for all the days. It is worth to mention that all the p-values were smaller than 0.001.

\FloatBarrier

\subsection{Robustness analysis}

To analyze the robustness of the studied models, we ran calibrations on 200 bootstrapped samples as described in Subsection \ref{s:meth:calib:robAn} and examined errors and the variation in the prices and coefficients.

For the initial points for the bootcalibrations, we chose the parameters estimated by the overall calibration (Table~\ref{t:overall:results}) while keeping the other calibration procedure parameters the same as before.

First, we examine the errors of prices and its variation with respect to the changing option structure. Then, we analyze the variation in the model parameter estimates. Finally, we summarize the results in a table and quantitatively compare the studied models to the Heston, Bates, and AFSVJD models earlier analyzed by \cite{PospisilSobotkaZiegler19ee}.

\subsubsection{Prices – errors and variation}

We examined how the model prices obtained from bootcalibrations differ from the market prices by calculating bootstrap relative errors $BRE_i$  using \eqref{e:meth:calib:robAn:BRE} and the variation price estimates by calculating the variances of absolute relative errors $V_i$ using \eqref{e:meth:calib:robAn:Variance_V_i}.

\begin{figure}[ht]
\centering
\includegraphics[width=.45\textwidth]{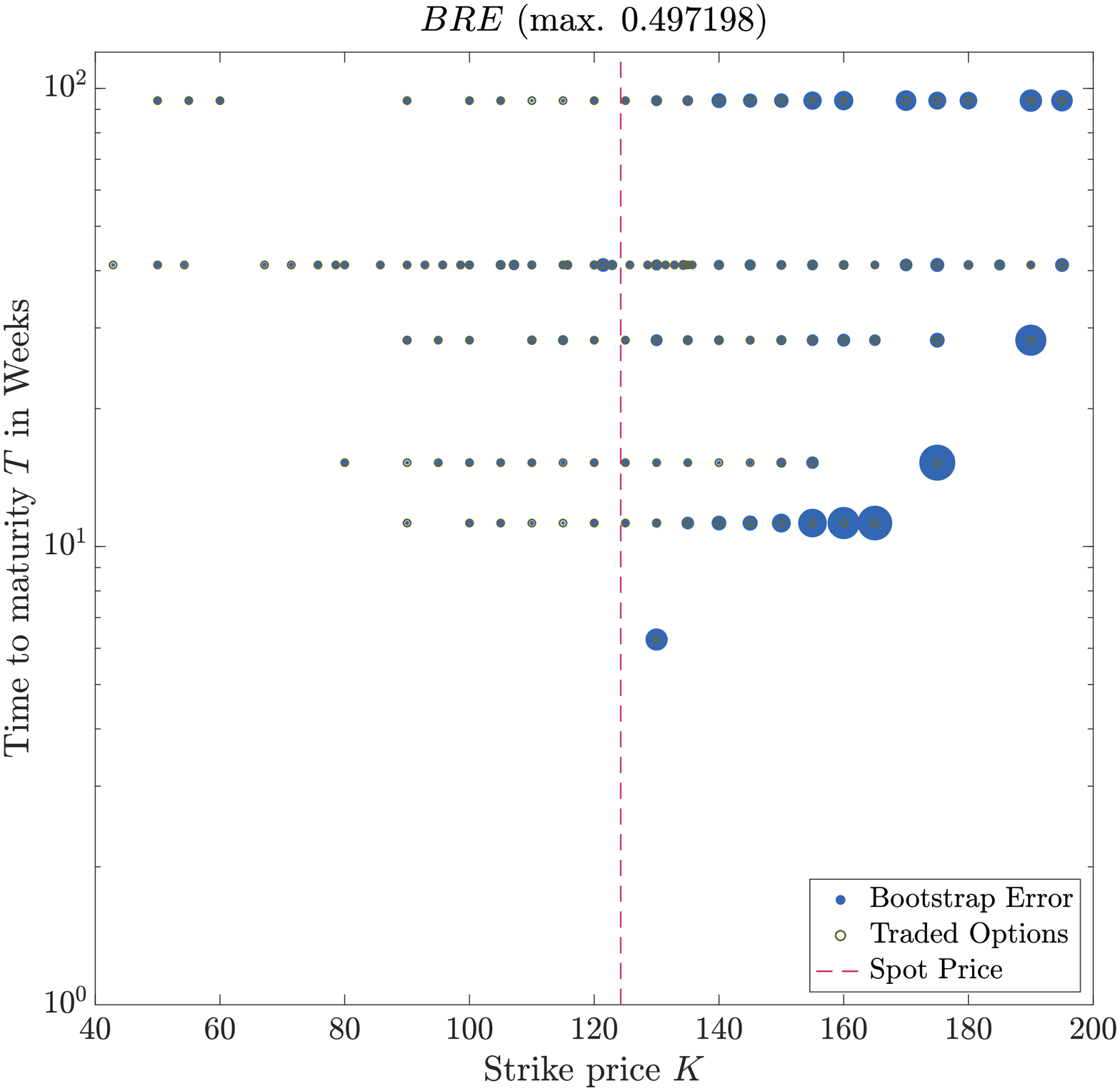}
\includegraphics[width=.45\textwidth]{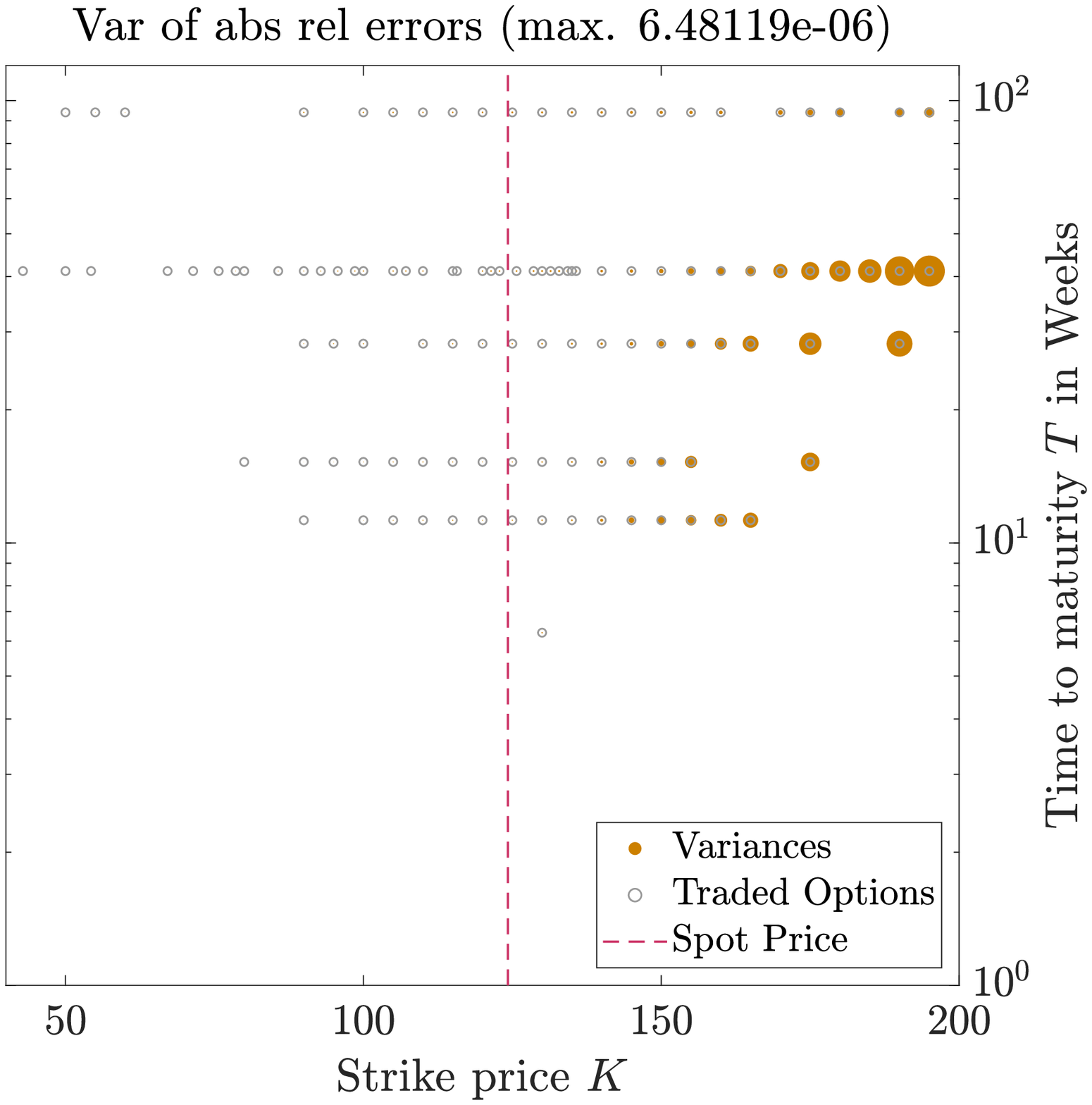}\\
\includegraphics[width=.45\textwidth]{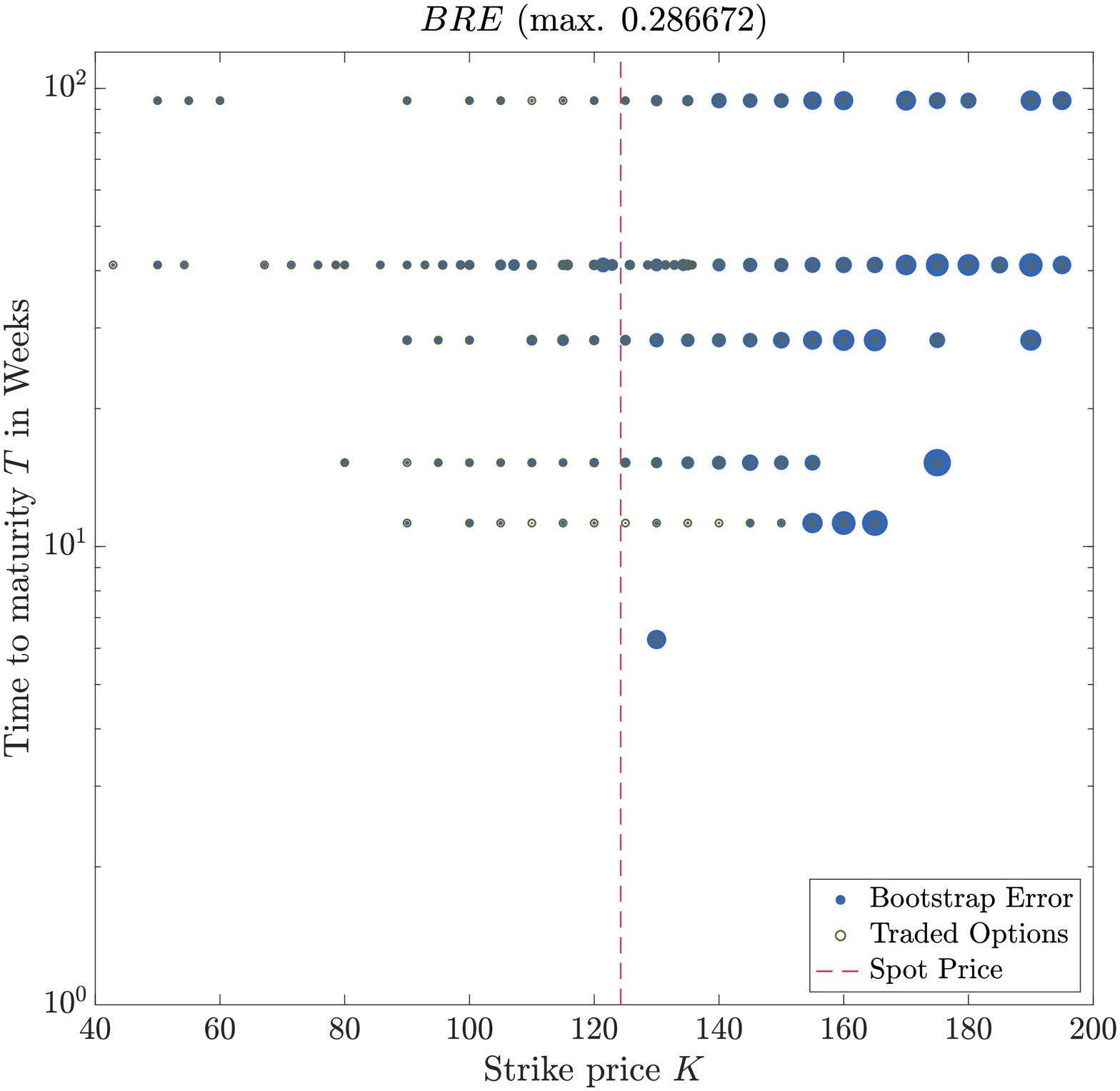}
\includegraphics[width=.45\textwidth]{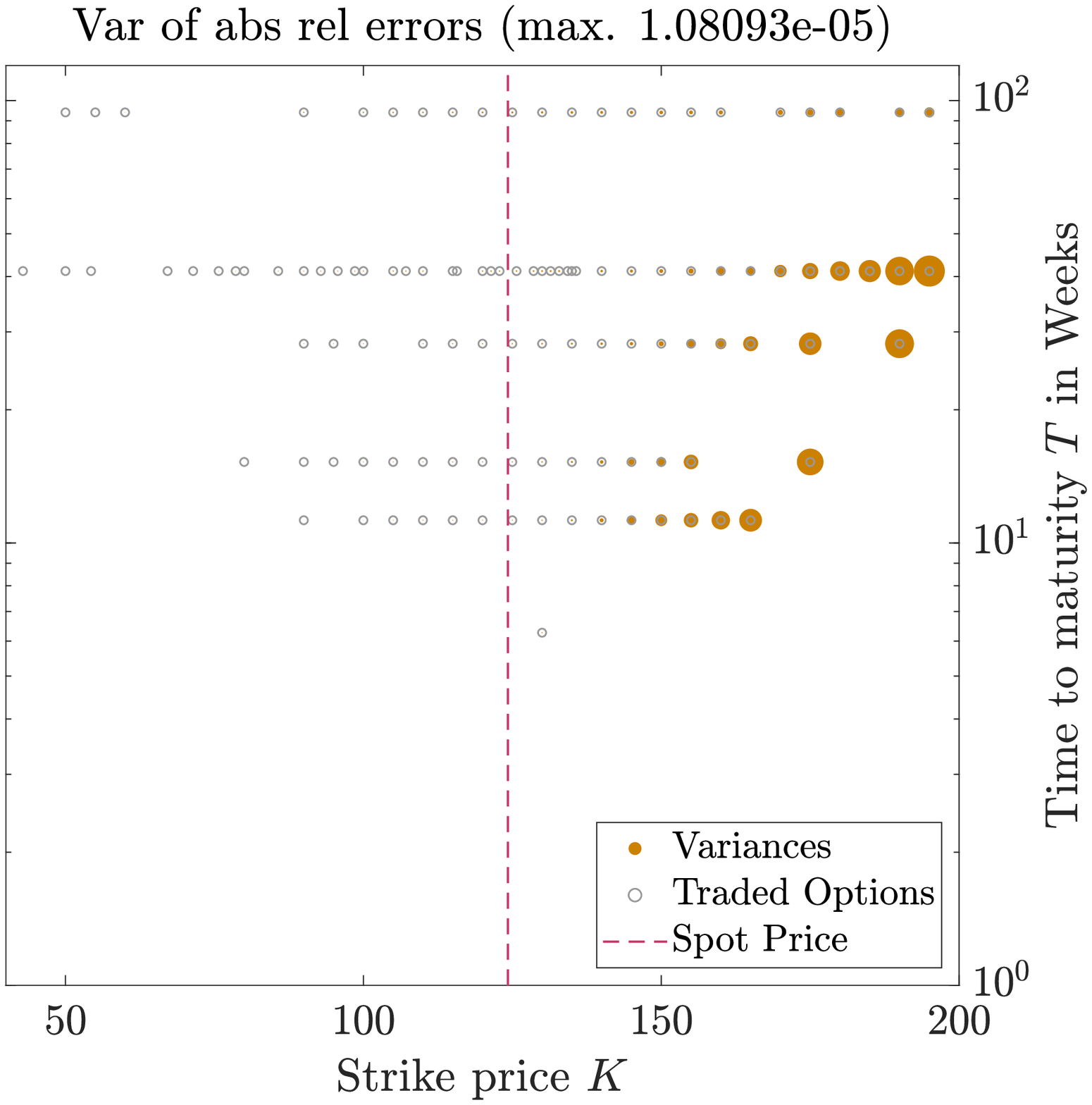}
\caption{Option data structure from April 1, where the diameter of the size of a disk depicts the bootstrap relative error $BRE_i$ \eqref{e:meth:calib:robAn:BRE} (left) and the variance $V_i$ \eqref{e:meth:calib:robAn:Variance_V_i} (right) corresponding to an option of a given combination of $K$ (x-axis) and $T$ (y-axis). The top row belongs to the RFSV model and the bottom one to rBergomi. \label{f:robAn:5-01:BRE_Var:fixed}}
\end{figure}

Figure \ref{f:robAn:5-01:BRE_Var:fixed} depicts the values of $BRE_i$ and $V_i$ for the option structure on April 1 produced by the $\alpha$RFSV model (top row) and the rBergomi model (bottom row). For both models, the largest errors and variation of errors concentrates on the right side relative to the spot price. This is expected since deep OTM options have zero intrinsic value and hence all the value comes from the time component associated with the probability of profitable exercise at expiration which is naturally more difficult to model. Comparing the results of two models on this particular day, rBergomi provides much better fit than $\alpha$RFSV. The variation in rBergomi errors is also smaller which indicates that the model is less sensitive to the changes in the option structure.

Complete results for all the three models for each day are available in similar format in Appendix of the thesis \cite{Matas21}. %% BLIND
By comparing all the figures, we saw the same pattern: larger errors for the OTM option. However, an interesting result is that the $\alpha$RFSV model has the smallest variation (sometimes by even more than two orders of magnitude) of the errors for all the days which suggests that the $\alpha$RFSV model may the most robust of the three models.

\FloatBarrier
\subsubsection{Variability of estimations of model coefficients}

In order to analyze the variability of the coefficient estimates obtained from the bootcalibrations, we plot and examine scatterplot matrices. 

\begin{figure}[ht]
\centering
\includegraphics[angle=90,width=\textwidth]{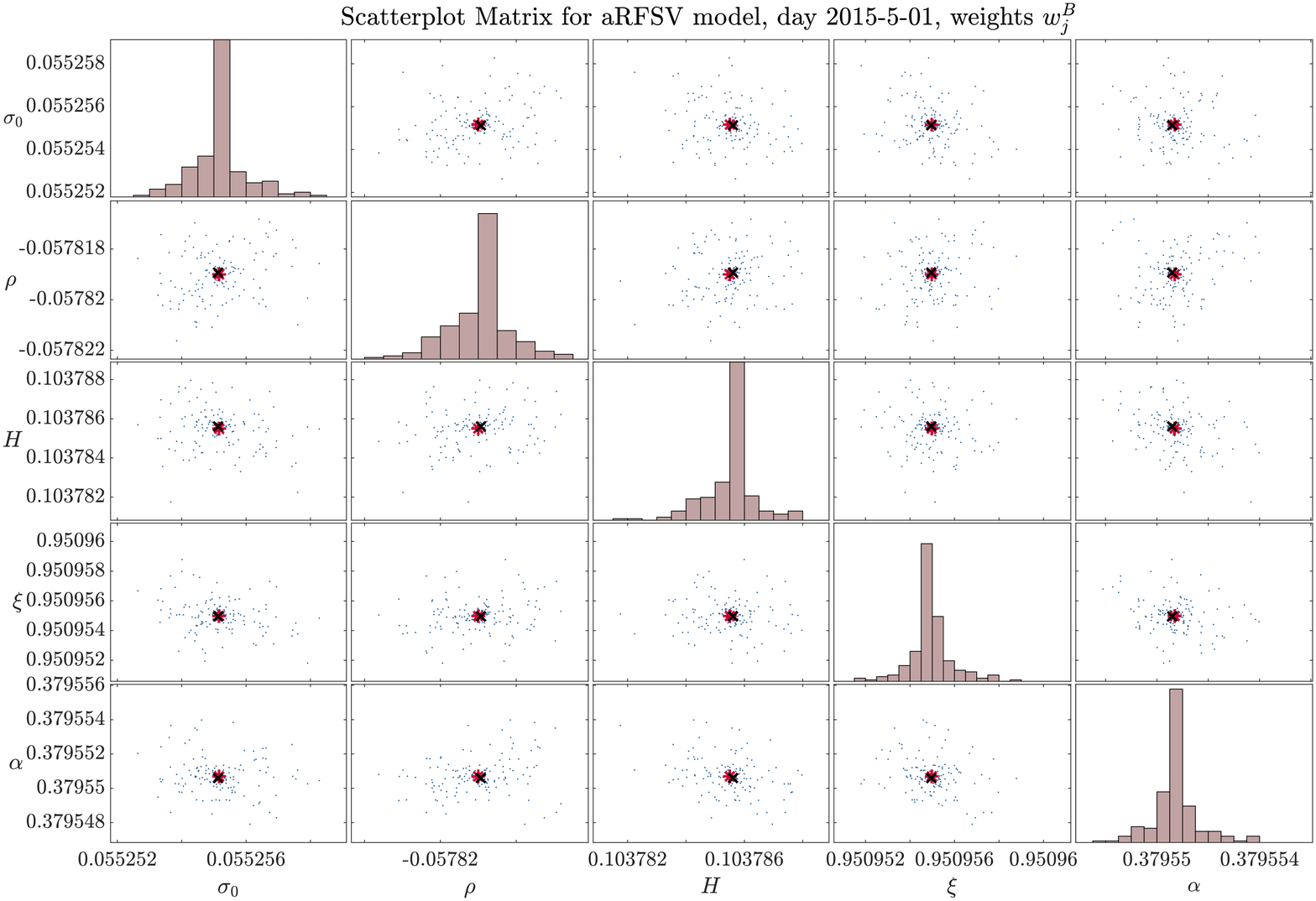}
\caption{A scatter plot matrix of the parameter estimates from bootcalibrations of the $\alpha$RFSV model for May 1. The red stars represent the bootstrap estimates \eqref{e:meth:calib:boot_mean_coeff}, while the black crosses represent the estimates from the overall calibration. \label{f:bootstrap:aRFSV:scatter:5-01}}
\end{figure}

Figure \ref{f:bootstrap:aRFSV:scatter:5-01} illustrates the model coefficient estimates of the $\alpha$RFSV model obtained from the bootcalibrations. Since the 5-dimensional parameter space is visualized as a matrix of 2D scatter plots, we can visually examine any patterns between the parameter estimates, while the histograms on the diagonal can provide some insight on the distributions of the parameter estimates. We can observe that there are no visible patterns and the distributions are symmetric with positive kurtosis which are both good properties for estimates. Also, notice that the variation around the bootstrap estimate is of a very small order of magnitude.

In fact, the scatter plot matrix in Figure \ref{f:bootstrap:aRFSV:scatter:5-01} is qualitatively identical to the scatter plot matrices for other models on any day. There are no visible patterns suggesting dependency between any two coefficient estimates, all the distribution are symmetric with positive kurtosis, and the variation around bootstrap estimates is remarkably low, especially compared to similar scatter plot matrices for the Heston, Bates, and AFSVJD models presented by \cite{PospisilSobotkaZiegler19ee}.

\subsubsection{Summary of robustness analysis and comparison to SV models}

Finally, to quantitatively compare the robustness analysis results of the three studied models to the Heston, Bates, and AFSVJD, we organized the results into Table \ref{t:summary:comparison:05-15} that provides results from May 15, but results from the remaining three days are qualitatively very similar and area vailable on request. The tablessummarize the variation both in absolute relative errors of prices and in the coefficient estimates. To quantify the variance in coefficient estimates by a single number for each model, we calculated the relative (normalized by average) inter-quartile ranges (IQRs) for each model coefficient from the 150 bootcalibrations and then calculated average and maximum from the relative IQRs. The variation in both AREs and coefficient estimates is smaller for rough models by several orders of magnitude compared to the non-rough SV models. Based on these results, we conclude that the rough models are more robust than Bates, Heston, and AFSVJD models.

\begin{table}[ht!]
\centering
\caption{Comparison of robustness analysis for different SV models; dataset 05/15.}
\label{t:summary:comparison:05-15}
\begin{tabular}{@{}l|rrr|rr@{}}
\toprule
         & \multicolumn{3}{c|}{BootARE} & \multicolumn{2}{c}{Coefficient Estimates} \\ %\midrule
Model    & Range  & IQR   & Std   & Rel IQR Avg & Rel IQR Max \\ \midrule
Bates    & 13.663 & 0.484 & 2.342 & 0.1193      & 0.2188      \\
AFSVJD   & 8.447  & 0.989 & 1.579 & 0.0746      & 0.1692      \\
Heston   & 14.194 & 0.583 & 2.377 & 0.0975      & 0.1856      \\
aRFSV    & 0.009  & 0.002 & 0.002 & 7.61E-06    & 2.17E-05    \\
rBergomi & 0.031  & 0.002 & 0.004 & 2.99E-05    & 5.32E-05    \\
RFSW     & 0.013  & 0.003 & 0.003 & 3.08E-06    & 9.50E-06    \\ \bottomrule
\end{tabular}
\end{table}

\subsection{Sensitivity analysis}

We conducted the sensitivity analysis as described in Subsection \ref{s:meth:calib:sensAn}, i.e., for each parameter in a given day and for a given model, we tested the null hypothesis that the distribution of the parameter estimates corresponding to the 3/8 "worst" bootcalibrations is the same as the distribution of the parameter estimates belonging to the 3/8 "best" bootcalibrations, using the KS test.

The KS test did not reject the null hypothesis for any of the parameter-model-day combination. That indicates that the studied models are not sensitive to changes in the option structure when being calibrated. Although considerable variation in the values of the $\alpha$RFSV is still prevalent, the results of the sensitivity analysis suggest that the variation comes mainly from the changes in the option structure, independently of the parameter estimates.

% ------------------------------------------------------------------------------ end: results.tex
% ------------------------------------------------------------------------------ begin: conclusion.tex
\section{Conclusion}\label{sec:conclusion}

{Initially, we compared the goodness of fit among the rough SV models we examined, as well as with the Heston, Bates, and AFSVJD models, using the average absolute relative error as the measure. Our findings indicated that none of the fractional SV models demonstrated superior performance for the used data sets. However, the $\alpha$RFSV model displayed the most consistent results. Notably, when RFSV exhibited better performance for a particular data set, the $\alpha$ parameter tended to be closer to 0. Conversely, when the rBergomi model provided a better fit, the $\alpha$ parameter was closer to 1. Nevertheless, our comparison of the average absolute relative error indicated that the rough models we investigated did not exhibit superiority over the Heston, Bates, and AFSVJD models.

Then we presented the parameter estimates of the overall calibrations and tested the parameters $H$ and $\alpha$ for significance. The two-sample t-test confirmed that both the parameters are very statistically significantly when calibrated for all the four data sets used.

Next, we analyzed the robustness of the rough SV models based on plots of BRE, variances of absolute relative errors across the bootstrapped data sets, and the scatter plot matrices of the parameter estimates. While the BREs were higher for the OTM option in all cases, an interesting results was that the $\alpha$RFSV had the smallest variation (sometimes by more than two orders of magnitude) of the errors for all days compared to the two other studied models. The scatterplot matrices revealed that there are no patterns suggesting that any pair of parameter estimates would be dependent and the variance of the estimates turned out to be remarkably small, especially compared to the standard SV models. We also provided a table that summarizes the robustness analysis results by quantifying the variation in parameter estimates by several standard statistics for all the studied models (both rough and standard SV). Based on these results, we concluded that the rough models are much more robust than Bates, Heston, and AFSVJD models.

Lastly, we tested the sensitivity of the models to the changes in the option structure when being calibrated. We used a Monte Carlo filtering technique and the KS test. The statistical procedure did not show that the fit of a given model is significantly sensitive to the changes in the option structure. Consequently, we concluded that the persisting variability in the errors originates from changes in the option structure, regardless of the estimated parameters.

During the process of writing the paper, several additional questions and issues arose. Regarding calibration, we could estimate some of the model coefficients from time series, e.g., the Hurst parameter $H$ can be estimated by the method proposed in \cite{Gatheral18}, or the coefficient $\rho$ can be estimated as the correlation between stock price returns and realized volatility changes. We could then analyze the robustness of the models for such cases in a similar fashion. Another possibility is to try a different approach for the calibration itself. Instead of the deterministic gradient-based trust region approach of \texttt{lsqnonlin()}, we could employ a stochastic approximation approach or even the deep-learning method developed by \cite{Horvath2021deep}. 

In the paper by \cite{MerinoPospisilSobotkaSottinenVives21ijtaf}, an approximation of the option price in the $\alpha$RFSV model was derived and numerical experiments therein propose a promising hybrid calibration scheme which combines the approximation formula alongside MC simulations. Since the aim of this paper was to study the model as accurately as possible, we avoided the usage of approximation formula in our robustness and sensitivity analyses tests, however, repeating the same experiments with the usage of approximation formula should be straightforward.

% ------------------------------------------------------------------------------ end: conclusion.tex
% ------------------------------------------------------------------------------ begin: acknowledgement.tex
\section*{Funding}
The work was partially supported by the Czech Science Foundation (GA\v{C}R) grant no. GA18-16680S ``Rough models of fractional stochastic volatility''.

\section*{Acknowledgements}
This work is a part of the Master's thesis \cite{Matas21} titled \emph{Rough fractional stochastic volatility models} that was written by Jan Matas and supervised by Jan Posp\'{\i}\v{s}il. 

Our sincere gratitude goes to Tom\'{a}\v{s} Sobotka for his valuable suggestions and insightful criticism and to all anonymous referees for their valuable comments and extensive suggestions.
Computational and storage resources were provided by the e-INFRA CZ project (ID:90254), supported by the Ministry of Education, Youth and Sports of the Czech Republic.

% ------------------------------------------------------------------------------ end: acknowledgement.tex

% ------------------------------------------------------------------------------ begin: references-export.tex

% ------------------------------------------------------------------------------ end: references-export.tex

\end{document}